\pdfoutput=1
\documentclass[preprint2,numberedappendix,iop]{emulateapj-rtx4}
\usepackage{graphicx,graphics,amsmath}
\usepackage{natbib}
\usepackage{bm,url}
\usepackage{times}
\usepackage{amssymb}
\usepackage{color}
\usepackage{float}
\usepackage{arydshln}
\newcommand{\approptoinn}[2]{\mathrel{\vcenter{
	\offinterlineskip\halign{\hfil$##$\cr
	#1\propto\cr\noalign{\kern2pt}#1\sim\cr\noalign{\kern-2pt}}}}}

\usepackage{color}

\def\bl{Babcock--Leighton}

\newcommand{\Fig}[1]{Figure~\ref{#1}}
\newcommand{\Tab}[1]{Table~\ref{#1}}

\newcommand{\mps}{m~s$^{-1}$}
\newcommand{\kmps}{km~s$^{-1}$}
\newcommand{\cmss}{cm$^2$~s$^{-1}$}
\newcommand{\vpCZ}{$\gamma_{\rm CZ}$}
\newcommand{\vpS}{$\gamma_{\rm S}$}

\begin{document}
\title{Recovery from Maunder-like Grand Minima in a Babcock--Leighton Solar Dynamo Model}
\medskip
\author{Bidya Binay Karak$^{1,2,3}$}
\email{karak.app@iitbhu.ac.in}
\author{Mark Miesch$^{3}$}
\email{miesch@ucar.edu}
\affiliation{$^{1}$Department of Physics, Indian Institute of Technology (Banaras Hindu University), Varanasi, India\\$^2$Indian Institute of Astrophysics, Koramangala, Bangalore 560034, India\\$^{3}$High Altitude Observatory, National Center for Atmospheric Research, 3080 Center Green Dr., Boulder, CO 80301, USA}
\date{\today}

\begin{abstract}
The Sun occasionally goes through Maunder-like extended grand minima when its magnetic activity drops
considerably from the normal activity level for several decades. Many possible theories
have been proposed to explain the origin of these minima.
However, how the Sun managed to recover from such inactive phases every time is even more enigmatic.
The Babcock--Leighton type dynamos, which are successful in explaining many 
features of the solar cycle remarkably well, 
are not expected to operate during grand minima
due to the lack of a sufficient number of sunspots.
In this Letter, we explore the question of how the Sun could recover from grand minima through the Babcock--Leighton dynamo. 
In our three-dimensional dynamo model, grand minima are produced spontaneously as a result of random variations in the tilt angle of emerging active regions.
We find that the Babcock-Leighton process can still operate during grand minima with only a minimal number of sunspots and 
that the model can emerge from such phases without the need for an additional generation mechanism for the poloidal field.
The essential ingredient in our model is a downward magnetic pumping which inhibits the diffusion of the magnetic flux across the solar surface.
\end{abstract}
\maketitle

\section{Introduction}
\label{sec:int}

The global magnetic field of Sun oscillates with polarity reversals every 11~years.
This oscillation is well reflected by the number of sunspots observed on the solar surface
and thus it is known as the sunspot cycle
or the solar cycle.
The solar cycle, however, is not regular. There was a time in the 17th century
when the sunspot number, as well as other proxies of solar activity e.g., the auroral occurrence, went to an unexpectedly low value 
for about 70~years.
This is the well-known   
Maunder minimum \citep{Eddy}. From indirect terrestrial proxies of solar activity,
we now know that this Maunder minimum is not unique and the Sun had many such events
with different durations in the past \citep{Beer,sol04,USK07}.
The interesting fact is that every time the Sun manages to recover to the normal magnetic activity from these grand minima. 
In fact, we now know that the magnetic field during the Maunder minimum was oscillating, 
implying that the underlying process of magnetic field generation was still occurring during the grand minima \citep{BTW98,Miy04}.

It is believed that a magnetohydrodynamics dynamo process, operating in the solar convection zone (CZ), is responsible for producing the solar magnetic cycle.
At present, the \bl\ type flux transport dynamo model is a popular paradigm for the solar cycle because of its success
in reproducing observations \citep{Cha10,Kar14a}.
In this model, the decay and dispersal of tilted bipolar magnetic regions (BMRs) near the solar surface produce the poloidal field---the \bl\ process \citep{Ba61,Le64}. 
The poloidal field is then transported to the bulk of the CZ through the turbulent diffusion and meridional circulation, where the winding of this field by differential rotation generates a toroidal field.
This model is constructed with an assumption that the toroidal flux near the base of the CZ (BCZ)
produces BMRs at the surface. The observed tilt of BMRs relative to an east-west orientation 
is attributed to Coriolis force during the rise of the toroidal flux through the CZ \citep[e.g.,][]{DC93}.

The BMR tilt is crucial in producing the poloidal field
in this model \citep{Das10}. While in observations the tilt systematically increases with latitude---Joy's law, there is a considerable scatter around this systematic variation \citep{SK12,Wang15,Arlt16}.
This scatter in the tilt angle causes a variation in the polar field \citep[][hereafter KM17]{JCS14,JCS15,HCM17,KM17}.
Based on this idea previous authors have included a random component in the
Babcock--Leighton source of their flux transport dynamo models and
have shown
that this random component can diminish the poloidal source and trigger a grand minimum \citep{Cha04,CK09,CK12,KC13,OK13,Ha14,Pas14,IAR17}.  However, these models do not explicitly take into account realistic BMR properties such as tilt angle scatter, flux distribution, and cycle-dependent emergence rate.  Furthermore, they do not fully explain how a Babcock-Leighton model can emerge from a grand minimum without some additional source of poloidal field such as a turbulent $\alpha$-effect.
Recently, \citet{LC17} (hereafter LC17) have developed a 2D$\times$2D coupled surface flux transport and flux transport dynamo model
in which the actual BMRs with observed properties have been included.
They have demonstrated that explicit tilt angle fluctuations can indeed induce grand minima.

In a newly developed 3D dynamo model (KM17), we have included the tilt angle fluctuations explicitly
and we have shown that the observed tilt scatter is capable of triggering grand minima events.
When using the currently observed Gaussian fluctuations with $\sigma_\delta=15^\circ$, the occurrence of grand minima in the model is somewhat less frequent than that inferred from terrestrial proxies \citep{USK07}.  
However, a solar-like frequency is found when we double the scatter. 
Here we use the enhanced
  tilt-angle scatter of 30$^\circ$ (double the observed value of 15$^\circ$) 
in order to facilitate our analysis by producing more frequent grand minima.  
However, we emphasize that the mechanism we describe here for emerging from a grand minimum 
is not sensitive to the value used for the tilt-angle scatter; solar-like models 
emerge from grand minima in the same way.  The essential emergence mechanism 
we describe (enabled by magnetic pumping) is also insensitive to other parameters of the model as well; 
one should be able to capture it even with a 2D model that does not have any explicit active regions. 

\begin{figure*} \centering
\includegraphics[width=2.4\columnwidth]{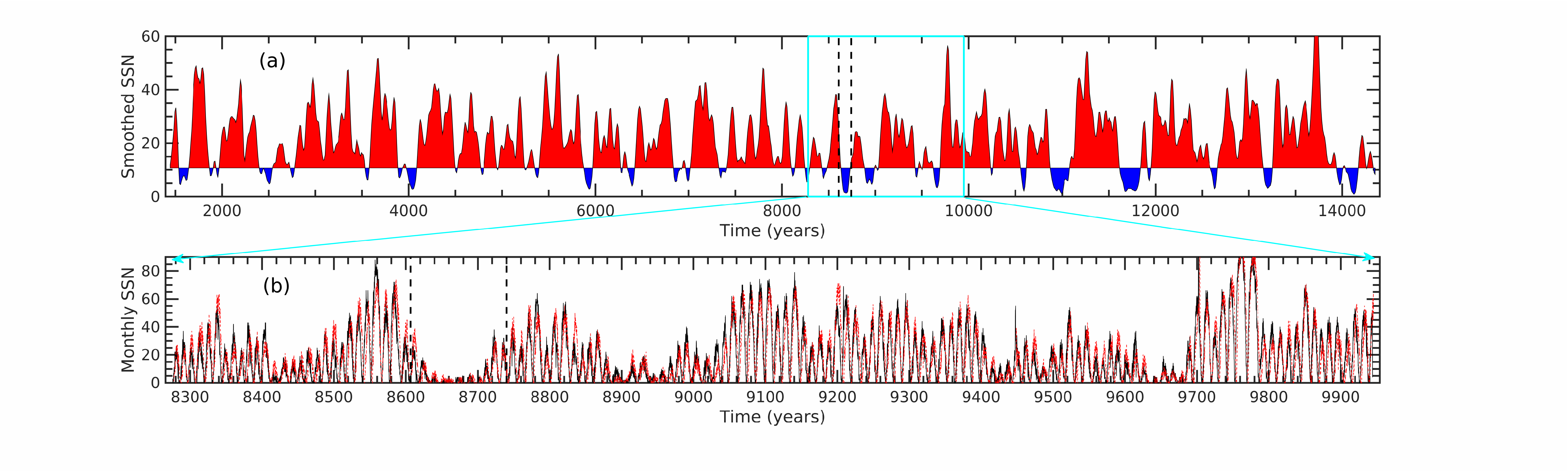}
\caption{(a) Temporal variation of the smoothed SSN from 13,000-year simulation of Run~B11. 
Blue shaded regions below the horizontal line 
represent the grand minima.
(b) Monthly smoothed (with boxcar average of 3-month width) 
SSN (black/red: north/south) shown only for a selected 1600-year interval.
Vertical dashed lines indicate the time-window which is focused in \Fig{fig:grandmin}.
}
\label{fig:30s}
\end{figure*}

Although previous studies demonstrate that tilt angle scatter can cause grand minima,
they do not explain the recovery of the Sun from such phases.
As BMRs are the only source for the generation of the poloidal field in the \bl\ type dynamos,
the generation of poloidal field becomes negligible during grand minima due to a fewer number of BMRs.
Thus the \bl\ dynamo may stop operating during grand minima and the Sun may not recover.
Previous studies suggested that an additional poloidal source (e.g., convective $\alpha$) 
is needed in order for the Sun to recover from grand minima \citep{KC13,Ha14,Pas14}. 
Indeed, LC17 observed
that when their model enters into an extended grand minimum of the weaker magnetic field,
the generation of poloidal field stops due to lack of BMRs and the model never recovers from that quiescent phase.
Their model recovers only when the magnetic field does not fall below a certain level.

\section{Model}
\label{sec:mod}
In this Letter, we explore the \bl\ dynamo mechanism 
during grand minima, focusing in particular on how the dynamo might recover from such episodes through the Babcock-Leighton process alone.
To do so, we first produce grand minima. 
In our study, we build on our recent work, KM17 which is an updated version of the original model \citep{MD14,MT16}.
In this model, BMRs are produced near the surface based on the toroidal flux at the BCZ
and most of the statistical properties of BMRs are based on solar observations.
We refer the readers to Section~2 of KM17 for the details of this model.
From KM17, we consider Runs~B10--11 in which the diffusivity in the bulk of the CZ is
in the order of $10^{12}$~\cmss.  
The BMR flux distribution is fixed at the observed value
and the time delay between two successive BMR emergences is obtained from 
the observed log-normal distribution 
which changes in response to the toroidal field at the base of the CZ
to allow more frequent BMRs when the toroidal
field is strong and vice versa.
The BMR tilt has a Gaussian scatter around Joy's law 
with standard deviation ($\sigma_\delta$) of $30^\circ$ ($15^\circ$) 
for Run B11 (B10). 
Another key ingredient of the model is the downward magnetic pumping 
which has the same form as in KM17 and it is given by
\begin{eqnarray}
\bm \gamma = - \frac{ \gamma_{\rm CZ} }{2}\left[1 + \mathrm{erf} \left(\frac{r - 0.725R_\odot}{0.01R_\odot}\right) \right]\hat{\mbox{\boldmath $r$}} \nonumber  \\
              - \frac{(\gamma_{\rm S} - \gamma_{\rm CZ})} {2}\left[1 + \mathrm{erf} \left(\frac{r - 0.9R_\odot}{0.02R_\odot}\right) \right]\hat{\mbox{\boldmath $r$}},
\label{eqpump}
\end{eqnarray}
where $\gamma_{\rm CZ} = 2$~\mps\ and $\gamma_{\rm S} = 20$~\mps.
Thus, the pumping $\bm \gamma$ has a value of $\gamma_{\rm S}$ only near the surface,
while in the rest of the CZ it is $\gamma_{\rm CZ}$.

Pumping is a process in which the magnetic flux can be transported 
in a stratified convective medium due to the topological asymmetry in the convective flow. 
A variety of theoretical and numerical models suggest that 
it is operating in the solar CZ, particularly near the surface
where both the velocity amplitude and the density stratification are largest
\citep[e.g.,][]{DY74,KR80,Tob98,PS93}.
Although at present, we do not have a stringent constraint on the strength of the pumping,
previous studies \citep{Kap06,Kar14b} suggest its value to be at least a few tenths of the convective velocity.
Keeping in mind that the upper layer of the Sun is highly convective with the observed surface convection
speed of a few \kmps, a value of $20$~\mps\ pumping speed is realistic \citep{Spr97,NSA09}.  
This value is also large enough to make the magnetic field near the surface approximately 
vertical, which improves the agreement between Babcock-Leighton dynamo models and 
Surface Flux Transport (SFT) models \footnote{\citet{Ca12} showed that
  the inclusion of magnetic pumping in Babcock-Leighton dynamo models 
can improve their agreement with SFT models.
  The radially downward magnetic pumping makes the field more radial near the surface and
  suppresses the vertical diffusion of the magnetic flux. To do this the radial advection
  of the field through downward pumping has to dominate over the diffusion.  
They showed that this requires $5 l/ \gamma_{\rm S} = l^2 / \eta$,
where $l = 0.1 R_\odot$ the depth of the near-surface layer where pumping is operating,
and $\eta$ is diffusion coefficient in the near-surface layer.
Taking $\eta = 3 \times 10^{12}$~cm$^2$~s$^{-1}$,
this gives the pumping speed $\gamma_{\rm S}  = 21.5$~\mps.
Thus the value of $\gamma_{\rm S}$ used in our simulations is in agreement
with this argument.}.

\section{Results and Discussion}
\label{sec:res}
A time series of sunspot number (SSN) obtained from 
Run~B11
is shown in \Fig{fig:30s}(a);
see \Tab{table1} for parameters
and KM17 for details of how the sunspots are produced in this model based on the toroidal field in the interior.
We note that this SSN is smoothed using the same procedure as done in \citet{USK07}, 
that is, we first bin the monthly SSN in 10-year intervals and then filter the data 
by averaging over five neighboring points using the Gleissberg's low-pass 
filter 1-2-2-2-1 \citep{Glei44}. 
The blue shading areas indicate the grand minima which are defined when the smoothed SSN goes below $50\%$ of the mean for at least two consecutive decades
(the same procedure as given in \citet{USK07}). 
To display the variation of the original SSN, including its 11-year periodicity, 
we show the monthly SSN variation for about 1600~years in \Fig{fig:30s}(b).
In \Fig{fig:30s}(a), we notice several grand minima; see Run B11 in \Tab{table1}. 
Durations of some of these grand minima are similar to the Maunder minimum and some are even longer.

\begin{figure*}
\centering
\includegraphics[width=1.6\columnwidth]{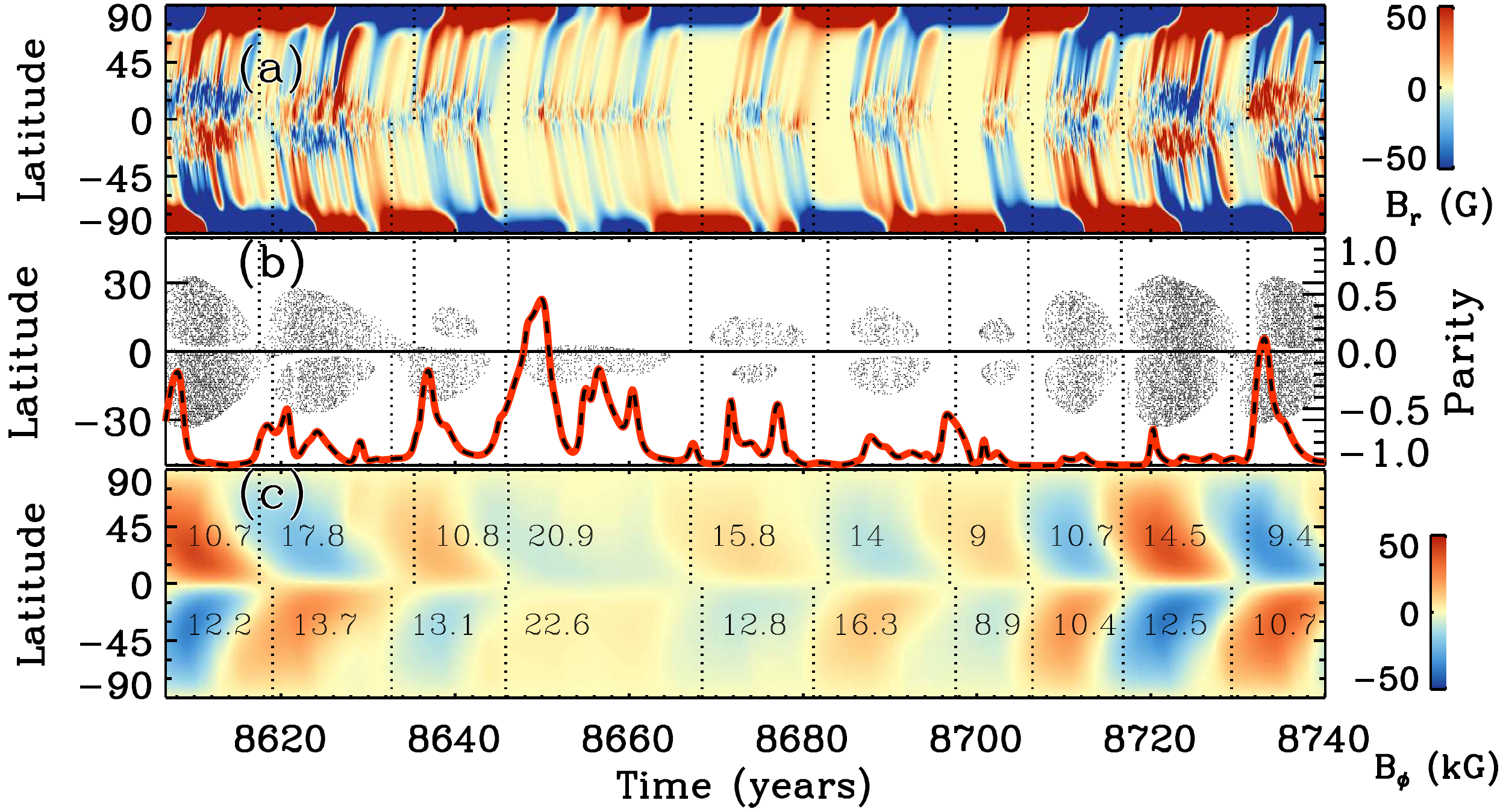}
\caption{Zoomed-in view of a grand minimum from \Fig{fig:30s}. 
Latitude-time variations of (a) surface radial field, (b) sunspots, and (c)
toroidal field at the BCZ.
The red/dashed line in (b) shows the parity of the toroidal  
field at the BCZ (computed 
by cross-correlating the field between two hemispheres
using Eq.~11 of KM17).
We note that for perfect anti-symmetric (symmetric) toroidal field,
the parity is expected to be $-1$ ($1$). 
Dotted lines mark 
the minima of sunspot cycles 
and periods in year are labeled in (c). 
The extrema of $B_r$ and $B_\phi$ are $[-1561, 1645]$~G and $[-38,40]$~kG, respectively.}
\label{fig:grandmin}
\end{figure*}

\begin{table}
\caption{Summary of simulations.
Parameters of Runs~B10--B11 are the same as in KM17,
while for Run~B2$^*$ $\Phi_0$ and $\sigma_{\delta}$ are different than the ones in KM17.
Run~B2$^*$ failed to recover after it entered into a grand minimum.
$T_{\rm sim}$, and $f_{\rm GM}$ denote the length of simulation,
and $\%$ of time spent in grand minima (GM), respectively.
}
\begin{center}
\begin{tabular}{lcccccc}
\hline
Run &$\Phi_0$& \vpCZ, \vpS (\mps) & $\sigma_{\delta}$&$T_{\rm sim}$ (yr) &$\#$ of GM& $f_{\rm GM}$\\
\hline
B10 & 2.4    & 2, 20~~~~~~~   & 15$^{\circ}$& 11650~~ & 17     & 11\%     \\
B11 & 2.4    & 2, 20~~~~~~~   & 30$^{\circ}$& 19214~~ & 46     & 17\%     \\
B2$^{*}$& 65 & 0, ~0~~~~~~~   & 30$^{\circ}$& ~~589~~ & ~1     & $\ldots$ \\
B14  & 2.4   & 2, 22~~~~~~~   & 30$^{\circ}$& ~2952~~ & ~5     & 17\%     \\
\hline
\end{tabular}
\end{center}
\label{table1}
\tablecomments{
The parameter $\Phi_0$ is used to boost the observed flux 
distribution in our model; see Equation~8 of KM17 for details.
}
\end{table}

To characterize the features of the grand minima produced in our
model, we highlight a few cycles from
8615--8740 years in \Fig{fig:grandmin}.  We notice that the period of
the first few cycles during this grand minimum is slightly longer than
the average period 
of 10.8 years.
This is consistent with the solar activity during Maunder minimum obtained 
from $^{14}$C data \citep{Miy04}, 
although $^{10}$Be data gives somewhat different picture \citep{FSB99,BTW98}.  
The longer cycle period
is expected when there are fewer BMRs at the beginning
of the grand minimum because with few BMRs the new poloidal flux
needs more time to accumulate and thus reverse the old polar flux.  
We further note
that during grand minima, BMRs appear near the equator which 
is consistent with the observational findings during the Maunder
minimum \citep{RN93}. 
Low-latitude BMRs appearance in our model is a consequence of 
 chosen latitude-dependent threshold field strength for BMR production; see KM17 for details.
Another distinct feature of grand
minima is the hemispheric asymmetry. Around the year 8660 in \Fig{fig:grandmin}(b), 
most BMRs are produced in the southern hemisphere (also see the parity of the field which is linked to the hemispheric asymmetry; KM17). 
A strong hemispheric asymmetry was also 
observed during the Maunder minimum \citep{RN93}.  
All these features (longer period, BMRs emergence near equator, and hemispheric asymmetry) 
are not limited to this grand minimum shown in \Fig{fig:grandmin},
they are also observed in other grand minima.

\begin{figure}
\centering
\includegraphics[width=0.9\columnwidth]{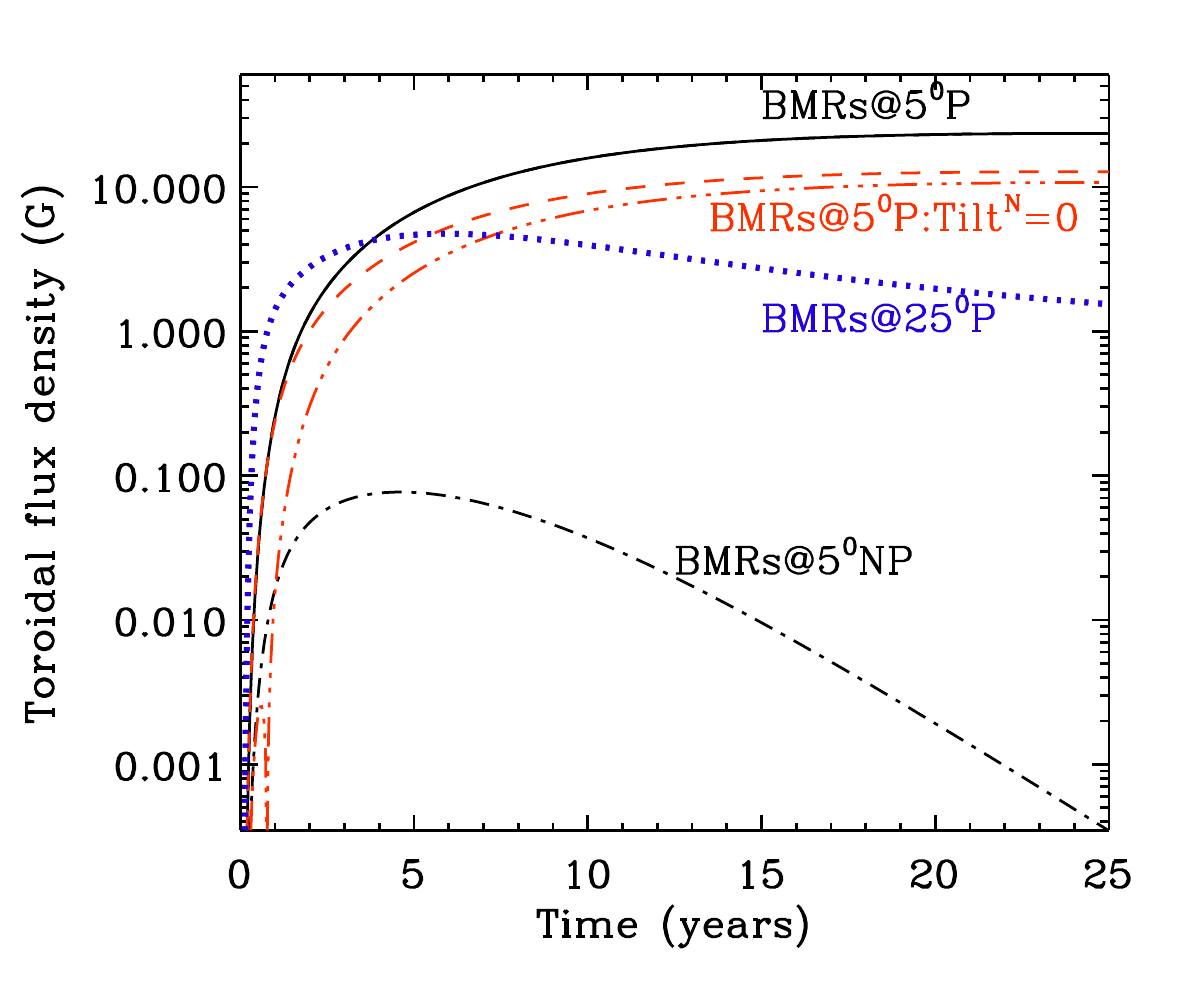}
\caption{Evolution of the absolute value of the toroidal flux density 
 obtained by averaging from $0^\circ$ to $30^\circ$ latitude at BCZ.
The solid and dotted lines
represent cases in which BMRs are deposited symmetrically at
$\pm5^\circ$ 
(BMRs@5$^\circ$P) 
and $\pm25^\circ$ latitudes 
(BMRs@25$^\circ$P),
respectively.  The red
lines (dash-double dotted for north and dashed for south) represent the case in
which BMRs are deposited at $\pm5^\circ$ but the northern BMR has
zero tilt 
(BMRs@5$^\circ$P:Tilt$^{\rm{N}}$=0).
The dash-dotted line represents the case in which BMRs are
deposited at $\pm5^\circ$ but the pumping is set to zero
(BMRs@5$^\circ$NP).
}
\label{fig:2BMRs} 
\end{figure}

In any \bl\ dynamo model, the only source of poloidal field is the tilted BMRs.
Thus, BMR emergence is essential to enable our model to emerge from grand minima.
In our model, the SSN during grand minima goes to a very small value but never becomes zero for much more than a year.
For example, the mean spot number during years 8650--8710 in \Fig{fig:30s}(b) is about $5.8$, which is only $13\%$ of the mean spot number from the entire simulation run.  
Thus the question is how those fewer sunspots during grand minima are capable of producing enough poloidal flux to maintain the dynamo?

It turns out that it is the downward magnetic pumping which enables our model to recover from grand minima even with a few sunspots.
The magnetic pumping near the surface makes the poloidal field radial and suppresses the diffusion of the horizontal field through the surface, as shown by \citet{Ca12,KC16}.
Thus when a few sunspots during grand minima produce poloidal flux, it remains in the CZ for many years. 
This poloidal flux
continuously produces toroidal flux through the $\Omega$ effect. The pumping
also does not allow this toroidal flux to diffuse through the solar surface. 
(The toroidal flux can diffuse across the equator but 
this diffusion can be balanced by its generation.)

To validate this idea, we examine the magnetic field generated
from the decay of two BMRs in this model.  We perform a
simulation by depositing one BMR at $5^\circ$
latitude and another at $-5^\circ$ latitude as an initial condition, with no other seed field present. 
Tilts of these BMRs are given by Joy's law with no scatter around it. The flux and other properties of these BMRs are identical.
The polar flux produced from the decay of these BMRs 
eventually produces toroidal flux near the BCZ as shown by the solid line 
(case: BMRs@5$^\circ$P)
in \Fig{fig:2BMRs}.
Now we repeat the same experiment by switching off the magnetic pumping.
The dash-dotted line in \Fig{fig:2BMRs} represents
this simulation
(case: BMRs@5$^\circ$NP).
We find that without pumping the toroidal flux becomes orders of magnitude smaller and decays indefinitely.
Within the context of grand minima, this implies that the poloidal flux produced by even a few BMRs will remain in the CZ long enough to be converted to toroidal flux through the differential rotation.  Eventually, the toroidal flux will become strong enough to trigger more BMR emergence, bringing the model out of the grand minimum.

We note that the recent 2D$\times$2D model of LC17, which
uses a much smaller tilt scatter 
than we have used in the present simulation,
shuts off entirely whenever it enters into a Maunder-like extended minimum.
While there are many fundamental differences between their model and ours, the major difference is 
that their model does not take into account magnetic pumping.  In their model, when
SSN falls below a certain level for a few years, the toroidal
field decays rapidly and once it falls below the threshold for
the spot production, no new spot can form.  This makes
the dynamo shut off completely.

\begin{figure}
\centering
\includegraphics[width=1.0\columnwidth]{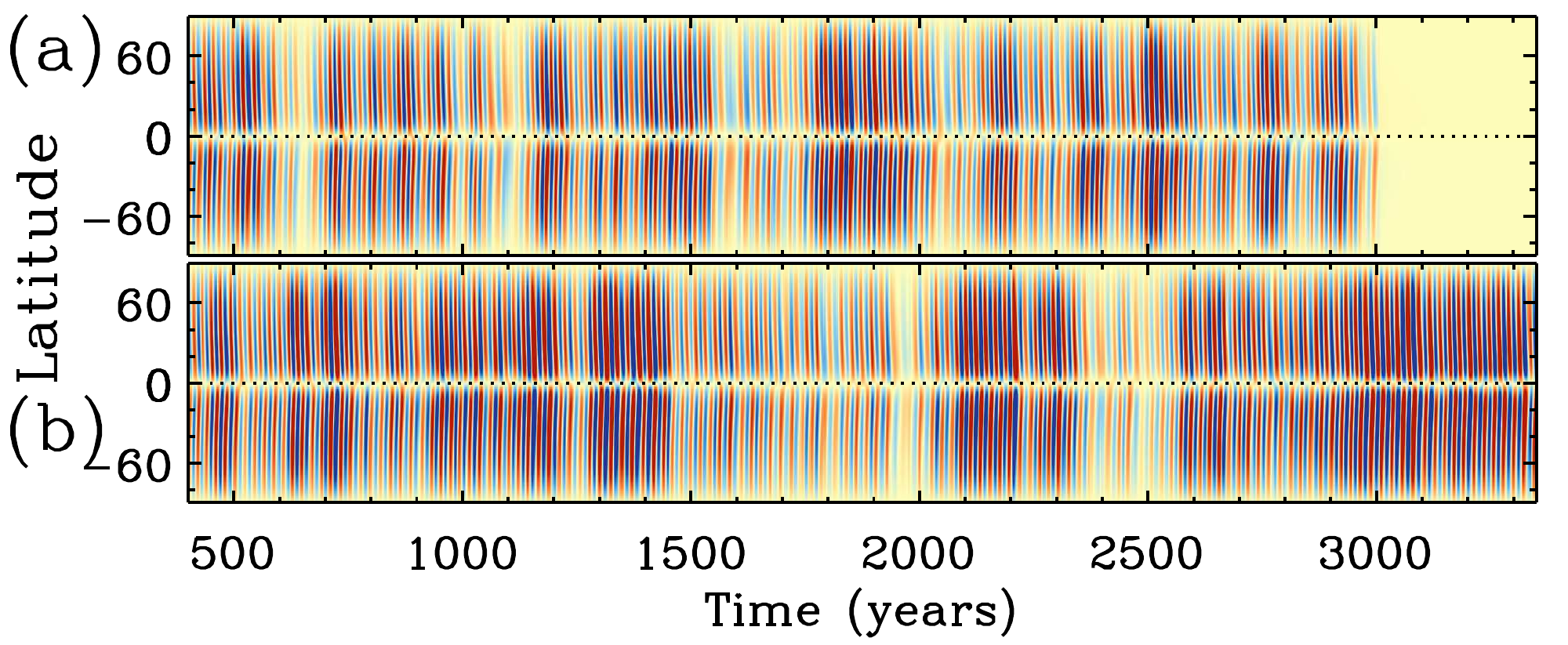}
\includegraphics[width=1.0\columnwidth]{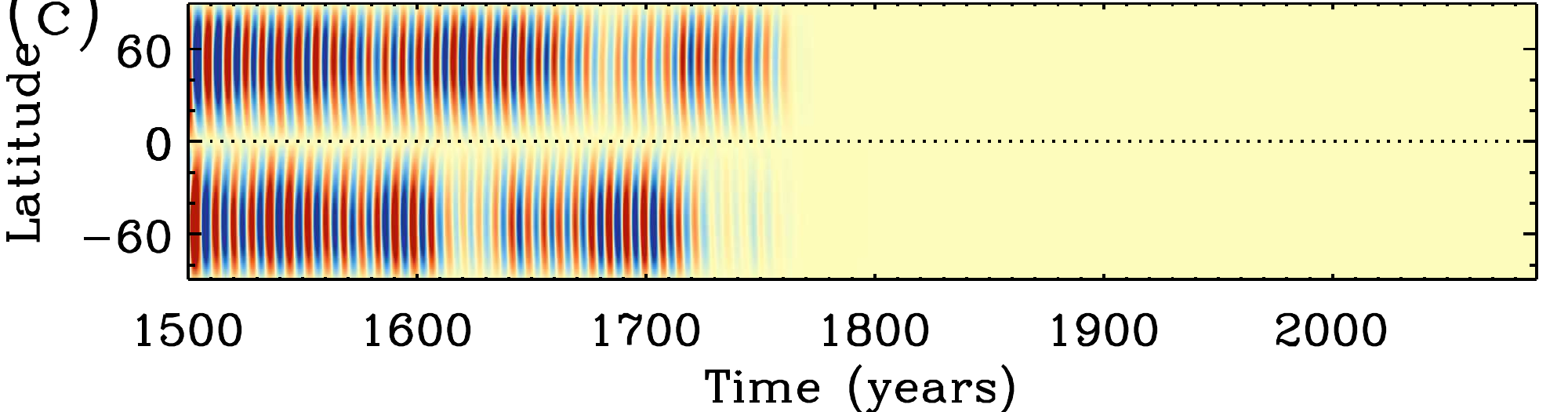}
\caption{Latitude-time variation of the toroidal field at the BCZ. 
(a) Demonstrates a case out of total 19,200 years of simulations (Run~B11) 
which could not recover from a grand minimum.
(b) Obtained from Run~B14 which is the same as (a) except higher pumping.
(c) From Run~B2$^*$ (without magnetic pumping).}
\label{fig:decay}
\end{figure}

To probe the above conclusion even further, we repeat our grand minima simulation without magnetic pumping (Run B2$^*$).
The tilt scatter and other parameters are the same as in Run~B11.
However, the spot flux (parameter $\Phi_0$ in the model) is increased to 65 from 2.4. 
This change is needed in order to make the dynamo supercritical since pumping enhances the dynamo efficiency \citep{KC16}.
The output of this simulation is shown in \Fig{fig:decay}(c). As expected, when the magnetic pumping is not included,
the model cannot recover from the grand minimum once it enters into it. 
Interestingly, when we restart this simulation 
with the snapshot right before it entered into the grand minimum ($t=1750$~years) as the initial condition 
but with magnetic pumping, then it recovers.

Another feature of our model that is beneficial for recovery from grand minima 
is the spontaneous emergence of BMRs at low latitudes, as evident in \Fig{fig:grandmin}(b).
These low-latitude spots are very efficient in generating poloidal flux in comparison to the higher latitude spots.
To demonstrate that this is true, we repeat the same simulation of two symmetric BMRs
as shown by the solid line in \Fig{fig:2BMRs}
(BMRs@5$^\circ$P),
 but instead of depositing them
at $\pm5^{\circ}$ latitudes, we deposit them at $\pm25^{\circ}$ latitudes. As usual, the tilt is obtained from Joy's law.
The dotted line in \Fig{fig:2BMRs} shows this simulation (note the log scale of the vertical axis); 
case: BMRs@25$^\circ$P.
On comparing with solid line, we confirm that
the BMR pair closer to the equator produces much larger toroidal
flux, although they have smaller tilt.  This is consistent with previous studies
\citep{JCS14, HCM17} which have shown that 
when BMR pairs emerge at low latitudes, the cancellation of flux across the equator is more efficient.  
Since this cancellation regulates the net flux in each hemisphere, it ultimately leads to stronger polar fields and, in turn, stronger toroidal fields.
 In our
model (and also in observations) BMRs during grand minima
are produced closer to the equator and these few 
low-latitude BMRs help the dynamo re-establish normal activity by enhancing the poloidal field generation.

Furthermore, our model has a strong hemispheric coupling (through
the turbulent diffusion). Due to this, if one hemisphere (for example, 
the northern hemisphere around 8660 years in \Fig{fig:grandmin}(b)) does
not get many BMRs, the other hemisphere
can supply some poloidal flux.  Thus strong hemispheric coupling is also beneficial for the
dynamo to recover from grand minima.

The relatively large tilt scatter in our model ($\sigma_\delta = 30^\circ$) 
can have a particularly important influence during grand minima, when the number of BMRs is small.
However, if a BMR pair near the equator gets very different tilts than given by Joy's law,
then a significant polar flux may be produced, unless when 
both BMRs have zero tilts or the same tilts with the same polarity
(i.e., one Hale and other anti-Hale).  For example, when one pair in one hemisphere has
zero tilt and the other pair's tilt is given by Joy's law, then they
still, produce a significant polar flux; see the red lines (dash-double dot for north
and dashed for south) in \Fig{fig:2BMRs}
(case: BMRs@5$^\circ$P:Tilt$^{\rm{N}}$=0).

We may ask the question, `Can the tilt scatter ever be large enough to make the poloidal flux 
extremely weak and the toroidal flux remains below the threshold for several years to produce no new spot?'
If this happens, then the dynamo may fail to recover from a grand minimum.
To explore this, we initiate different realizations of Run~B11 by using different random seeds 
for the tilt angle scatter, the time delay, and the BMR flux distribution.
In about 19,200~years of total simulation time, we found a case in which the model failed to recover from a grand minimum and
the dynamo shut off completely; see \Fig{fig:decay}(a).
In this case, the model could not recover
because SSN went to a very low value for many years and the poloidal field
generated from those few spots could not overcome the diffusion of the fields. However, the most interesting fact is that when we repeat this simulation 
with the same initial condition and same realizations of fluctuations 
but with an increased magnetic pumping \vpS\ 
of 22~\mps\ (instead of 20~\mps, which is the value for Runs~B10--11),
we do not get any dying dynamo; see Run~B14 in \Tab{table1} and \Fig{fig:decay}(b).
This slight increase in the magnetic pumping is enough to enable the dynamo to recover from all grand minima as discussed above.
 
\section{Conclusion}
\label{sec:con}
We find that the \bl\ process can still
operate during grand minima even with a few spots.
This result was unexpected and it strikingly contrasts to previous studies \citep{KC13,Ha14,Pas14,LC17},
which suggested that the \bl\ process cannot operate during Maunder-like
minima.  The \bl\ dynamo, of course, cannot operate when there are no sunspots
and that can happen if the magnetic field during grand minima goes to a very small value (below the threshold for spot generation).
However, we expect that this is not happening in Sun because
during grand
minima, at least during the Maunder minimum, there were still some
sunspots \citep{Vaq16,ZP16}.

We have demonstrated that magnetic pumping can sustain a Babcock-Leighton dynamo throughout a grand minimum and enable it to re-establish normal activity.  
It achieves this by suppressing diffusive losses, allowing toroidal magnetic flux to accumulate and amplify until it is large enough to produce sunspots (BMRs).    
The few sunspots during the grand minimum are enough to sustain the cycle, in part because they tend to emerge at low latitudes, which maximizes the efficiency of poloidal flux generation.  In contrast to other Babcock-Leighton models, there is no need to invoke an alternative source of poloidal field such as a turbulent $\alpha$-effect.  The sporadic appearance of sunspots at low latitudes in the model, often with substantial north-south asymmetry, is reminiscent of sunspot observations during the Maunder Minimum.   Our results therefore suggest that the Babcock-Leighton mechanism may have been sufficient to sustain the solar cycle throughout the Maunder Minimum and into its subsequent recovery, with similar implications for previous grand minima. 

\begin{acknowledgements}
We thank the anonymous referee, Mausumi Dikpati, Ricky Egeland, and Lisa Upton for reading this manuscript and providing valuable comments.
We also thank Robert Cameron and Dibyendu Nandi for past inspiring discussions related to this study.
BBK is supported by the NASA Living With a Star Jack Eddy Postdoctoral Fellowship Program,
administered by the University Corporation for Atmospheric Research.
The National Center for Atmospheric Research is sponsored by the National Science Foundation.
Computations were carried out with resources provided by NASA's High-End Computing program (Pleiades) and by NCAR (Yellowstone).
\end{acknowledgements}

\bibliographystyle{apj}
\bibliography{paper}

\begin{thebibliography}{45}
\expandafter\ifx\csname natexlab\endcsname\relax\def\natexlab#1{#1}\fi

\bibitem[{{Arlt} {et~al.}(2016){Arlt}, {Senthamizh Pavai}, {Schmiel}, \&
  {Spada}}]{Arlt16}
{Arlt}, R., {Senthamizh Pavai}, V., {Schmiel}, C., \& {Spada}, F. 2016, \aap,
  595, A104

\bibitem[{{Babcock}(1961)}]{Ba61}
{Babcock}, H.~W. 1961, \apj, 133, 572

\bibitem[{{Beer} {et~al.}(1990){Beer}, {Blinov}, {Bonani}, {Hofmann}, \&
  {Finkel}}]{Beer}
{Beer}, J., {Blinov}, A., {Bonani}, G., {Hofmann}, H.~J., \& {Finkel}, R.~C.
  1990, \nat, 347, 164

\bibitem[{{Beer} {et~al.}(1998){Beer}, {Tobias}, \& {Weiss}}]{BTW98}
{Beer}, J., {Tobias}, S., \& {Weiss}, N. 1998, \solphys, 181, 237

\bibitem[{{Cameron} {et~al.}(2012){Cameron}, {Schmitt}, {Jiang}, \& {I{\c
  s}{\i}k}}]{Ca12}
{Cameron}, R.~H., {Schmitt}, D., {Jiang}, J., \& {I{\c s}{\i}k}, E. 2012, \aap,
  542, A127

\bibitem[{{Charbonneau}(2010)}]{Cha10}
{Charbonneau}, P. 2010, Liv. Rev. Sol. Phys., 7, 3

\bibitem[{{Charbonneau} {et~al.}(2004){Charbonneau}, {Blais-Laurier}, \&
  {St-Jean}}]{Cha04}
{Charbonneau}, P., {Blais-Laurier}, G., \& {St-Jean}, C. 2004, \apjl, 616, L183

\bibitem[{{Choudhuri} \& {Karak}(2009)}]{CK09}
{Choudhuri}, A.~R., \& {Karak}, B.~B. 2009, Res. Astron. Astrophys., 9, 953

\bibitem[{{Choudhuri} \& {Karak}(2012)}]{CK12}
---. 2012, Phys. Rev. Lett., 109, 171103

\bibitem[{{Dasi-Espuig} {et~al.}(2010){Dasi-Espuig}, {Solanki}, {Krivova},
  {Cameron}, \& {Pe{\~n}uela}}]{Das10}
{Dasi-Espuig}, M., {Solanki}, S.~K., {Krivova}, N.~A., {Cameron}, R., \&
  {Pe{\~n}uela}, T. 2010, \aap, 518, A7

\bibitem[{{Drobyshevski} \& {Yuferev}(1974)}]{DY74}
{Drobyshevski}, E.~M., \& {Yuferev}, V.~S. 1974, Journal of Fluid Mechanics,
  65, 33

\bibitem[{{D'Silva} \& {Choudhuri}(1993)}]{DC93}
{D'Silva}, S., \& {Choudhuri}, A.~R. 1993, \aap, 272, 621

\bibitem[{{Eddy}(1976)}]{Eddy}
{Eddy}, J.~A. 1976, Science, 192, 1189

\bibitem[{{Fligge} {et~al.}(1999){Fligge}, {Solanki}, \& {Beer}}]{FSB99}
{Fligge}, M., {Solanki}, S.~K., \& {Beer}, J. 1999, \aap, 346, 313

\bibitem[{{Gleissberg}(1944)}]{Glei44}
{Gleissberg}, W. 1944, Terrestrial Magnetism and Atmospheric Electricity
  (Journal of Geophysical Research), 49, 243

\bibitem[{{Hazra} {et~al.}(2017){Hazra}, {Choudhuri}, \& {Miesch}}]{HCM17}
{Hazra}, G., {Choudhuri}, A.~R., \& {Miesch}, M.~S. 2017, \apj, 835, 39

\bibitem[{{Hazra} {et~al.}(2014){Hazra}, {Passos}, \& {Nandy}}]{Ha14}
{Hazra}, S., {Passos}, D., \& {Nandy}, D. 2014, \apj, 789, 5

\bibitem[{{Inceoglu} {et~al.}(2017){Inceoglu}, {Arlt}, \& {Rempel}}]{IAR17}
{Inceoglu}, F., {Arlt}, R., \& {Rempel}, M. 2017, \apj, 848, 93

\bibitem[{{Jiang} {et~al.}(2014){Jiang}, {Cameron}, \& {Sch{\"u}ssler}}]{JCS14}
{Jiang}, J., {Cameron}, R.~H., \& {Sch{\"u}ssler}, M. 2014, \apj, 791, 5

\bibitem[{{Jiang} {et~al.}(2015){Jiang}, {Cameron}, \& {Sch{\"u}ssler}}]{JCS15}
---. 2015, \apjl, 808, L28

\bibitem[{{K{\"a}pyl{\"a}} {et~al.}(2006){K{\"a}pyl{\"a}}, {Korpi},
  {Ossendrijver}, \& {Stix}}]{Kap06}
{K{\"a}pyl{\"a}}, P.~J., {Korpi}, M.~J., {Ossendrijver}, M., \& {Stix}, M.
  2006, \aap, 455, 401

\bibitem[{{Karak} \& {Cameron}(2016)}]{KC16}
{Karak}, B.~B., \& {Cameron}, R. 2016, \apj, 832, 94

\bibitem[{{Karak} \& {Choudhuri}(2013)}]{KC13}
{Karak}, B.~B., \& {Choudhuri}, A.~R. 2013, Res. Astron. Astrophys., 13, 1339

\bibitem[{{Karak} {et~al.}(2014{\natexlab{a}}){Karak}, {Jiang}, {Miesch},
  {Charbonneau}, \& {Choudhuri}}]{Kar14a}
{Karak}, B.~B., {Jiang}, J., {Miesch}, M.~S., {Charbonneau}, P., \&
  {Choudhuri}, A.~R. 2014{\natexlab{a}}, \ssr, 186, 561

\bibitem[{{Karak} \& {Miesch}(2017)}]{KM17}
{Karak}, B.~B., \& {Miesch}, M. 2017, \apj, 847, 69

\bibitem[{{Karak} {et~al.}(2014{\natexlab{b}}){Karak}, {Rheinhardt},
  {Brandenburg}, {K{\"a}pyl{\"a}}, \& {K{\"a}pyl{\"a}}}]{Kar14b}
{Karak}, B.~B., {Rheinhardt}, M., {Brandenburg}, A., {K{\"a}pyl{\"a}}, P.~J.,
  \& {K{\"a}pyl{\"a}}, M.~J. 2014{\natexlab{b}}, \apj, 795, 16

\bibitem[{{Krause} \& {R{\"a}dler}(1980)}]{KR80}
{Krause}, F., \& {R{\"a}dler}, K.~H. 1980, {Mean-field magneto\-hydro\-dynamics
  and dynamo theory} (Oxford: Pergamon Press)

\bibitem[{{Leighton}(1964)}]{Le64}
{Leighton}, R.~B. 1964, \apj, 140, 1547

\bibitem[{{Lemerle} \& {Charbonneau}(2017)}]{LC17}
{Lemerle}, A., \& {Charbonneau}, P. 2017, \apj, 834, 133

\bibitem[{{Miesch} \& {Dikpati}(2014)}]{MD14}
{Miesch}, M.~S., \& {Dikpati}, M. 2014, \apjl, 785, L8

\bibitem[{{Miesch} \& {Teweldebirhan}(2016)}]{MT16}
{Miesch}, M.~S., \& {Teweldebirhan}, K. 2016, \ssr

\bibitem[{{Miyahara} {et~al.}(2004){Miyahara}, {Masuda}, {Muraki}, {Furuzawa},
  {Menjo}, \& {Nakamura}}]{Miy04}
{Miyahara}, H., {Masuda}, K., {Muraki}, Y., {Furuzawa}, H., {Menjo}, H., \&
  {Nakamura}, T. 2004, \solphys, 224, 317

\bibitem[{{Nordlund} {et~al.}(2009){Nordlund}, {Stein}, \& {Asplund}}]{NSA09}
{Nordlund}, {\AA}., {Stein}, R.~F., \& {Asplund}, M. 2009, Living Reviews in
  Solar Physics, 6, 2

\bibitem[{{Olemskoy} \& {Kitchatinov}(2013)}]{OK13}
{Olemskoy}, S.~V., \& {Kitchatinov}, L.~L. 2013, \apj, 777, 71

\bibitem[{{Passos} {et~al.}(2014){Passos}, {Nandy}, {Hazra}, \&
  {Lopes}}]{Pas14}
{Passos}, D., {Nandy}, D., {Hazra}, S., \& {Lopes}, I. 2014, \aap, 563, A18

\bibitem[{{Petrovay} \& {Szakaly}(1993)}]{PS93}
{Petrovay}, K., \& {Szakaly}, G. 1993, \aap, 274, 543

\bibitem[{{Ribes} \& {Nesme-Ribes}(1993)}]{RN93}
{Ribes}, J.~C., \& {Nesme-Ribes}, E. 1993, \aap, 276, 549

\bibitem[{{Solanki} {et~al.}(2004){Solanki}, {Usoskin}, {Kromer},
  {Sch{\"u}ssler}, \& {Beer}}]{sol04}
{Solanki}, S.~K., {Usoskin}, I.~G., {Kromer}, B., {Sch{\"u}ssler}, M., \&
  {Beer}, J. 2004, \nat, 431, 1084

\bibitem[{{Spruit}(1997)}]{Spr97}
{Spruit}, H. 1997, MmSAI, 68, 397

\bibitem[{{Stenflo} \& {Kosovichev}(2012)}]{SK12}
{Stenflo}, J.~O., \& {Kosovichev}, A.~G. 2012, \apj, 745, 129

\bibitem[{{Tobias} {et~al.}(1998){Tobias}, {Brummell}, {Clune}, \&
  {Toomre}}]{Tob98}
{Tobias}, S.~M., {Brummell}, N.~H., {Clune}, T.~L., \& {Toomre}, J. 1998,
  \apjl, 502, L177

\bibitem[{{Usoskin} {et~al.}(2007){Usoskin}, {Solanki}, \& {Kovaltsov}}]{USK07}
{Usoskin}, I.~G., {Solanki}, S.~K., \& {Kovaltsov}, G.~A. 2007, \aap, 471, 301

\bibitem[{{Vaquero} {et~al.}(2016){Vaquero}, {Svalgaard}, {Carrasco}, {Clette},
  {Lef{\`e}vre}, {Gallego}, {Arlt}, {Aparicio}, {Richard}, \& {Howe}}]{Vaq16}
{Vaquero}, J.~M., {et~al.} 2016, \solphys, 291, 3061

\bibitem[{{Wang} {et~al.}(2015){Wang}, {Colaninno}, {Baranyi}, \&
  {Li}}]{Wang15}
{Wang}, Y.-M., {Colaninno}, R.~C., {Baranyi}, T., \& {Li}, J. 2015, \apj, 798,
  50

\bibitem[{{Zolotova} \& {Ponyavin}(2016)}]{ZP16}
{Zolotova}, N.~V., \& {Ponyavin}, D.~I. 2016, \solphys, 291, 2869

\end{thebibliography}
\end{document}